# Limits To Certainty in QoS Pricing and Bandwidth

L. Jean Camp
jean_camp@harvard.edu
617/496-6331
Harvard University

Carolyn Gideon
carolyn@the-friedmans.org
781/842-0415
Harvard University

## 1. Introduction

Advanced services require more reliable bandwidth than currently provided by the Internet Protocol, even with the reliability enhancements provided by TCP. More reliable bandwidth will be provided through QoS (quality of service), as currently discussed widely. Yet QoS has some implications beyond providing ubiquitous access to advance Internet service, which are of interest from a policy perspective. In particular, what are the implications for price of Internet services, for those who choose QoS? Further, how will these changes impact demand and universal service for the Internet.

Certainty in price has been shown to be important to consumers in general (Anania, Solomon, 1995; Odlyzk, 2000) and to universal service in particular (Camp & Tsang, 2000; Schement & Mueller, 1996). Simultaneously, consumption of services requiring some certainty of bandwidth, e.g. high quality streaming video and audio, has been increasing. Usage-based pricing methods have seemed an optimal method for addressing the need for better or guaranteed service, (e.g., Mackie-Mason & Varian, 1995; Key and Massoulie, 1999; Bailey, Nagel & Raghavan), though they reduce users' ability to predict the prices they will pay and total costs of usage they will incur.

This paper explores the relationship between certainty of bandwidth and certainty of price for Internet services over a statistically shared network. Both are important policy objectives. Certainty of bandwidth is necessary if the Internet is to serve as a platform for advanced communications, information and entertainment services, or even for telephony. Certainty of price is important for universal service. More specifically, we illustrate that certainty of bandwidth and certainty of price are mutually exclusive in a statistically shared network (of which the Internet is the canonical example).

Section 2 of this paper outlines the importance of certainty of price and how it might be achieved. Section 3 explains the importance of certainty in bandwidth and how different QoS protocols impact variations in bandwidth. Here, we use standard queuing models to show that increasing certainty in bandwidth results in decreased certainty in price. Conversely, setting a constant price with cost-based pricing in a statistically shared network results in delay. This is followed by a discussion of market evidence for this inverse relationship between certainty of price and certainty of bandwidth. Finally, we present conclusions and implications to be considered by policy makers.

Our assumptions are that statistically shared networks can be modeled using standard queuing models, and that pricing is cost-based. The results in this paper are not novel; but rather well known. What is unique in this work is the observation that well-know network phenomena have this particular fundamental economic implication. The findings of this paper imply that premium priced products with guaranteed bandwidth (e.g., voice services) will continue to find a market. The excess capacity of such offerings is not efficient in the technical sense, but can be justified by understanding how the network may be shared by flows with different requirements. Further, this implies that when building virtual circuits or offering premium service in a statistically shared network, marginal cost based pricing will not be feasible.

## 2. Certainty of Price: Marginal Cost

In this section we describe marginal-cost based pricing in a best-effort network. We describe what conditions must be true for certainty of price to exist. We also describe the reasons we believe that certainty of price is important from a policy perspective.

*Best-effort network*

Currently users obtain Internet service with certainty of price. For example, ISPs now offer a simple monthly subscription price, or a price depending on connection time. However, this is done whether by dial-up modem, Ethernet, cable Ethernet, or DSL by multiplexing subscribers and providing them with either only statistical bandwidth guarantees or no bandwidth guarantees at all (e.g., best effort). While such subscription pricing schemes are designed to cover the fixed costs of the network, they do not cover the marginal cost of delay to other users.

In a best-effort network, the marginal cost of transmitting a packet is a function of the state of network congestion. When a packet enters a network that is not congested, its existence and use of network resources does not result in any form of cost to other users or owners/operators/administrators of the network. Since the resources are all in place and not needed for any other use, the marginal cost of transmitting this packet is zero. However, when a packet enters a network that is congested, the resources needed to transmit this packet will cause other users to experience an increase in delay, or congestion. The total cost of this incremental delay to all other users is the marginal cost of transmitting this packet:

$$MC = \Sigma\Delta\tau, \text{ over all packets,}$$

*where MC is marginal cost, and*
$\Delta\tau$ *is the change in delay caused by this packet.*

Without complete knowledge of the state of the network, this cost can get huge very quickly and unknowingly to the user. Think of a user sending an extremely small packet when the network is congested. If this packet causes an infinitesimal, even undetectable, increase in delay to several millions of users, the marginal cost is very large, as the summation of several million infinitesimal amounts. With marginal-cost pricing in a best-effort network, a user can unknowingly incur large charges that result in bills the customer may be unwilling or unable to pay. The customer does not know if the network is congested, or to what degree it is congested. There are many factors related to the way the network is being used at different points at each minute that determine the amount of incremental delay a new packet of any given size will cause. It is impossible for a user to attain knowledge of these factors when deciding to send a particular packet at any given moment. The user also does not know the amount of incremental delay its packet causes other users, and how many of those users there are. Thus there is no way for even a sophisticated budget-conscious user to calculate his cost of sending a packet. This has potential to greatly impair universal service, as such unexpected increases in the cost of using the Internet may cause users to drop from the network, as discussed below.

*Conditions for certainty of price*
When network prices are set at marginal cost, and marginal cost is a function of delay as described above, then the uncertainty of the state of congestion in the network at any given moment will result in uncertainty of price. Thus, certainty of price requires certainty of the state of the network. Conceptually, such certainty can be achieved in three ways: (1) access to continually instantaneously updated information of network conditions, (2) excessive resources to provide a constant state of excess capacity for any possible level of peak network usage, or (3) blocking mechanisms that prevent packets from entering the network if any component to be used is already at capacity.

In a best-effort network, when there is congestion, packets, even those dropped, use resources that slow the entire system further (because they must be resent and because TCP rapidly slows down the speed at which it sends things) and increase delay. Therefore, packets entering the network when it is congested increase delay of all other users and so have marginal cost greater than zero. If packets are automatically dropped when the network is congested, then only those packets that arrive when there is no congestion will be delivered. These packets have a marginal cost of zero. In this type of network, only packets with a marginal cost of zero are delivered. Any packet with a positive marginal cost will not be delivered and thus there is no additional price to be paid by the user. This happens in QoS networks, where packets can be blocked from entering the system when there is congestion, and packets from users who have paid a higher price for greater bandwidth availability (or to reserve specific bandwidth) are given a higher priority for entering the system. This is discussed below in the following section.

*Importance of certainty of price*
The existing literature on network pricing and subscriber demand supports the preference for price certainty by significant portions of the subscriber population. Studies of telephone service demand have shown that users prefer flat-rate prices to measured-rate prices (*cite?*). Further studies have shown that users that drop their telephone service, either voluntarily or through inability to pay their telephone bill, do so because the uncertainty of usage-based pricing for certain telephone services result in unexpectedly large bills that are unaffordable (Schement & Mueller, 1996). More recently, studies of Internet access demand have similarly found that subscribers often prefer certainty in price. In a market experiment of demand for Internet access (INDEX), subjects seemed to prefer flat pricing for access to usage-based pricing when the flat price was comparable to major ISP subscription prices for unlimited use. Flat prices were often chosen even when usage-based prices would have resulted in lower weekly cost to the user (Chu, 1999). Both the preference shown for fixed-price service and the evidence from telephone service that usage-based pricing has the potential to cause users to incur service costs beyond their means suggest that uncertainty in price for network information services will have adverse affects on universal service for these services. Price uncertainty is likely to cause many users to choose not receive information services either because they fear bills they cannot afford to pay or because it causes budgeting difficulty. Thus certainty of price is an important policy objective for both diffusion and universal service purposes.

Best-effort networks with flat rate pricing suffer from tragedy of the commons. As each user pursues his own best interest, trying to send packets unaware of the state of the network, he unknowingly imposes costs on other users, which though undetectable on a packet-by-packet basis to individual users, aggregate into a network that does not function well. Pricing mechanisms that allow for charging users for this delay caused to others, in the form of marginal cost pricing described above, will cause users to incur huge costs unexpectedly which then is likely to result in a financial inability for many to use the Internet, even though they can well afford service when used only at non congested times. Therefore, it is difficult to imagine a scenario where a best-effort network can meet objectives of functioning well technologically, efficiently allocating resources (i.e. bandwidth), and achieving ubiquitous deployment and adoption, either with fixed-rate or marginal-cost pricing. This is the motivation behind current efforts to develop QoS mechanisms into the Internet. The following section discusses these efforts and the implications for certainty of bandwidth.

## 3. Certainty of Bandwidth

The relationship between certainty of price and certainty of bandwidth can be illustrated using the most basic queuing models for intserv or diffserv. The inverse relationship between certainty of delay and bandwidth found here is well-know; however, the pricing implications of these observations have never been made explicit.

Predictability in quality means that certain resources, most critically bandwidth, are certain to be available. Strictly reserving resources over the Internet isn't currently possible, although certainly some of the networks connected to the Internet enable resource reservation. Predictability in quality would be necessary for high-quality video conference or broadcast-quality (NTSC) streaming video, or even long distance voice.

There are two fundamental mechanisms for QoS over IP-based networks intserv (Zhang, Deering, Estrin, Shenker & Zappala, 1993), and diffserv (Clark, 1996). Here we describe theses mechanisms and present simple queuing models for each case.

The concept of queuing treats the network as a black box that has a rate of service and a stream of packets arriving. A queue is described according to the distributions that best describe the arrival and service of the packets, as well as the number of servers. The queue is denoted x/y/n where x denotes the arrival distribution, y denotes the service distribution, and n is the number of servers. If the arrival is Poisson and the service is Poisson then the queue is called a M/M/1 queue where the first M indicates that the distribution of the arrivals is Poison, the second M indicates that the distribution of service is exponential, and 1 indicates that the system can be modeled as a single queue. Many different queues can be modeled as a single queue as long as the distribution of arrivals and service are independent and identically distributed. While this is a gross simplification it is a remarkably useful simplification that will provide further illustration of our hypothesis.

The trivial observation is this: the mean and variance of an expediential distribution are equal. Thus an increase in the delay and an increase in the uncertainty of the delay increase the marginal cost of sending the packet.

With no QoS guarantees the network can be modeled, roughly, as an M/M/1 queue with:

*mean arrival rate:* $\lambda$, *and*
*service rate:* $\mu$.

This yields an average delay in such a best-efforts network of

$$\tau = (1/\mu)/(1-\rho), \textit{ where } \rho=\lambda/\mu \quad 0\leq\rho\leq 1.$$

In a statistically shared network modeled as a M/M/1 queue the delay can be assumed to be Poisson. $\rho$ is the ration of the demand to resources available.

*Intserv*

Intserv is so called because the reservation of bandwidth is integrated into the network connection. Intserv is based on the Resource Reservation Protocol (RSVP). With intserv the requestor of a connection sends out a request for bandwidth reservation. The routers along the path agree to reserve bandwidth and a virtual circuit with guaranteed through-put is created for the duration of the service. At this simple conceptual level the mode also applies to frame relay services, although clearly there are important technical distinctions.

The most trivial way to model such an event would be to assume that there was one server with service rate $\mu$ and then consider this broken into two servers each with a distinct service rate, $\mu_1$ and $\mu_2$. This would err in that the two flows are also distinct; it is not simply the case that the same flow is now served by two queues. Instead consider the creation of two queues.

The first queue is the queue enabled by the virtual circuit. This can be considered the reserved resources queue, where there is greater certainty that the necessary bandwidth will be provided as resources are reserved as per intserv. The virtual circuit creates a queue $q_1$ from the original queue $q$ leaving the remaining resources and flow to create best-effort queue $q_2$. Since intserv by definition provides adequate service, we will propose that

$$\lambda_2 /\mu_2 > \lambda_1/\mu_1,$$

*where* $\lambda = \lambda_1 + \lambda_2$ *and* $\mu_2 = \mu - \mu_1$.
*i.e.*, $\rho_2 > \rho_1$
*i.e.*, $\tau_2 > \tau_1$

We compare the situation to the case in which there is no virtual circuit so that there is the assumption that addition traffic is not entering the network. That is, we assume the increased bandwidth does not encourage additional use.

The marginal cost of intserv can be seen in the increased delay faced by each packet entering the second queue. The increased delay is the difference between the delay of the original queue and the delay of the new, reduced services best-effort queue:

$$\Delta\tau = (\mu_1 - \lambda_1)/(\mu - \lambda)(\mu_2 - \lambda_2), \text{ or}$$
$$\Delta\tau = \mu_1(1-\rho_1)/\mu_2\mu(1-\rho)(1-\rho_2).$$

This increase in delay is the mean increase as seen by the packets entering the second queue. The mean is expected to increase with reserved bandwidth and the difference is the marginal cost associated with the creation of the intserv queue. The denominator is clearly the inverse of the delay experienced by the reserved resources queue. Obviously delay is never negative. Thus the increase in delay and therefore increase in variance in delay caused is always positive.

Since this can be approximated as a Poisson process, the mean is equal to the variance. Thus not only are there marginal costs in terms of the creation of the intserv queue, the uncertainty associated with the delay in the network increases along with the mean. Thus the creation of a reserved resource intserv queue increases the certainty of bandwidth for packets in this queue, since adequate bandwidth is provided, but decreases the certainty of the price paid for these packets as the variance of delay caused to the best-effort queue is increased. Simple observation illustrates that the lower the utilization of the intserv queue, meaning the higher the certainty of bandwidth, the higher the variance associated with the delay in the best effort queue which is the dominant component of the marginal cost. Thus not only is certainty of bandwidth and certainty of price mutually exclusive, they are inversely proportional.

Similarly, the second best-effort queue sees an increase in the uncertainty of bandwidth available, as there is less bandwidth to be distributed and some bandwidth is unused even in the face of demand, resulting in higher mean and variance in the delay they incur. However, the delay these packets can cause would decrease primarily because there now exist a subset of packets that can be delayed by a packet in the lower priority queue $q_2$, compared to the original single queue, $q$. Recall

$$MC = \Sigma\Delta\tau$$

over all packets in a best-effort network, and note that now

$$MC = \Sigma\Delta\tau_2$$

with the summation only over those packets not travelling in the reserved resource queue.

Thus while the delay experienced by the best-effort queue increases, the delay caused by these packets decreases. Therefore the certainty of cost increases as the certainty in bandwidth decreases.

*Diffserv*

Another mechanism for creating higher quality of service is diffserv. diffserv is so-called because instead of offering service which is integrated into the network service is offered by differentiating between packets. Packets are marked "in" or "out" based on the agreement between the subscriber and the network services provider. All packets of lower priority "out" are served after the packets of high priority.

For the original queue again the delay remains

$$\tau = (1/\mu)/(1-\rho), \textit{ where } \rho=\lambda/\mu.$$

However, now the case is one of a single queue with lower and higher priorities. The well documented result is that the mean and the variance of the delay incurred by the lower priority queue increases. Thus again the increased certainty in bandwidth experienced by the higher-priority queue has a corresponding increase in uncertainty in price in a marginal-cost-pricing regime.

Also notable is the fact that the packets in the lower-priority queue both incur more delay and cause less delay. Thus in a marginal cost based pricing scheme the certainty of bandwidth increases the uncertainty of cost.

These conclusions are based on the assumption of a Poisson distribution, where mean and variance are equal. However, Gupta et. al. (2001) find in their simulation model of the Internet that variance often exceeds the mean. This strengthens the above result, as the increased certainty of bandwidth provided by creation of the high-priority queue increases the uncertainty of delay even more than proportionally, as the uncertainty of the delay increases by more than the mean of the delay.

## 4. The Market for Certainty of Price

This section presents some empirical market evidence for the relationship between certainty of price and certainty of bandwidth found above.

The most dramatic examples of this can be found in DSL, business DSL, and frame relay prices. In DSL a subscriber has a dedicated line up to a DSLAM, usually at the switch, where multiple DLS lines are multiplexed together before being sent out on a shared packet-based line.

In the case of business DSL the provider offers a service level agreement (SLA) that identifies the minimum and maximum bandwidth provided. A SLA provides bounded uncertainty about bandwidth. Business DSL services provide a service level agreement that sets a minimum level of service.

Tn lines provide certainty using frame relay technologies. The bandwidth on a Tn line is guaranteed.

Consider a DSL line that has a speed of 1.5mbs. In contrast a cable modem line may have 1.5mbs but be shared by ten people. Thus both have the same possible maximum bandwidth but the DSL user has greater certainty. As DSL line prices and cable prices have converged, DSL providers have also increased the use DSL multiplexers, thus decreasing the certainty. In addition the use of rate adaptive DSL allows LECs to extend DSL further from the home office at the cost of providing lower and less certain bandwidth. This is one anecdotal market example of a decrease in the certainty of bandwidth coming with an increase in the certainty of price.

A T1 line provides guaranteed bandwidth via frame relay service, again approaching 1.5mbps. Frame relay is a packet-based WAN communications protocol that can be easily used for switched as well as routed service. Frame relay networks are often shown as clouds like the Internet cloud. However, frame relay connections are all made via virtual circuits. While the proverbial cloud of uncertain routing paths and service availability is seen by every packet, the cloud in frame relay networks is seen only during the establishment of virtual circuits. Once the virtual circuit is established the frames themselves face no such uncertainty.[1] A T1 line functions on the basis of a committed information rate, CIR. Data are marked as within the CIR or beyond the CIR and frames that are beyond the CIR are discarded first in the case of congestion. Thus frame relay uses both the preferential discard of packets of diffserv and the reservation of resources of RSVP.

Business DSL provides a commitment of some level of service on average, T1 lines provides the commitment of a minimal level of service. DSL provides a potential maximum level with no contractually set minimum.

Further analysis of the data is certainly needed, including the various incentives and examination of the specific service level agreements offered in the various categories. Such work is in progress.

---

[1] Certainly automatic rerouting takes place but this is explicit route reconfiguration for the entire virtual circuit rather than rerouting of an individual packet.

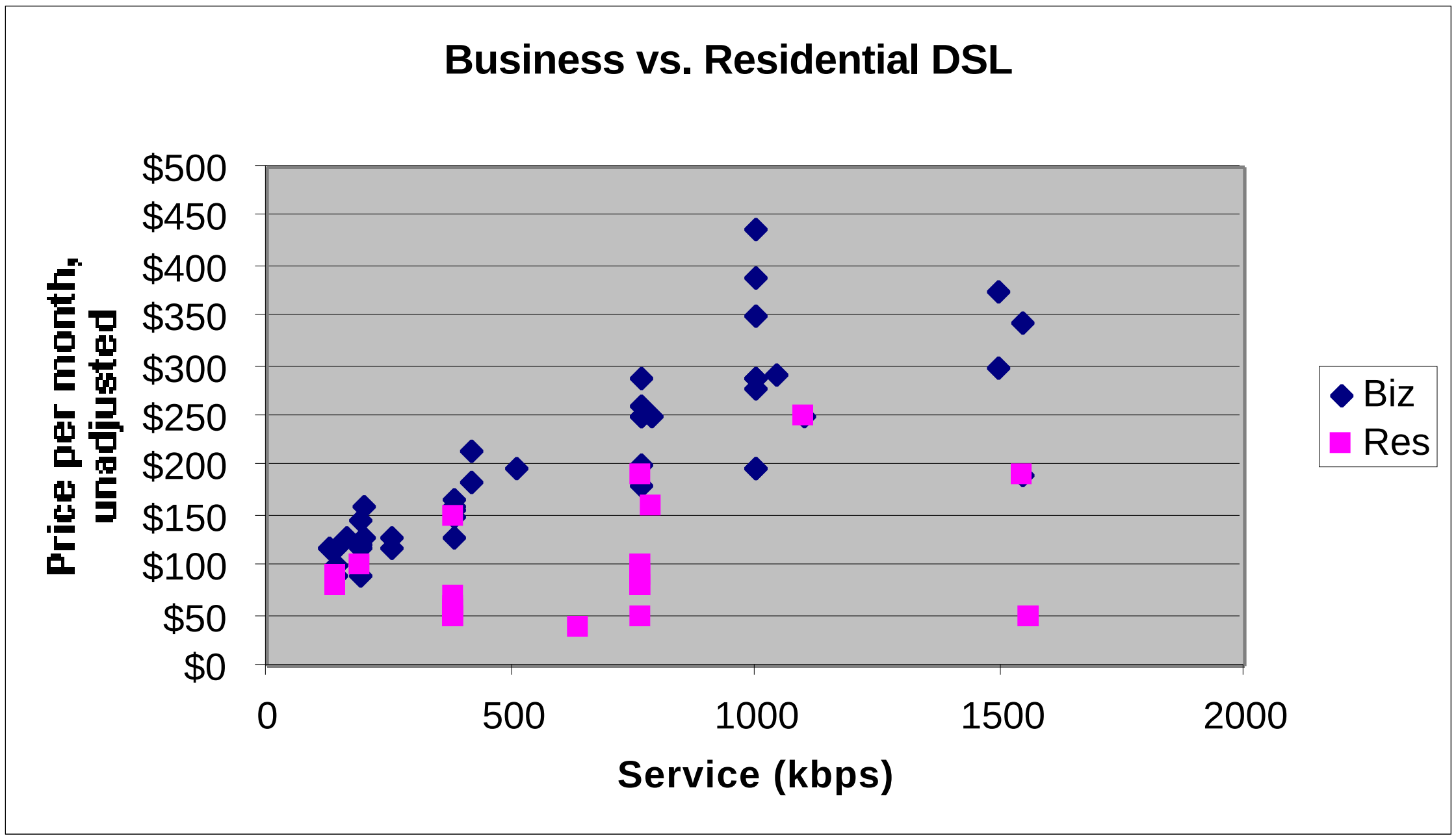

The chart above shows that residential DSL (■ Res) users who are not guaranteed bandwidth via SLAs pay less for the same amount of bandwidth than business users (◆ Bus). Business users have a higher certainty of bandwidth due to SLAs. In the weakest, this as least suggests a relationship between certainty and price. Certainly further consideration of this market would include comparisons of frame-relay services and T1 services. Simple observation illustrates that frame relay services are orders of magnitude more expensive then equivalent shared bandwidth services.

## 5. Conclusions

It is not possible to use marginal-cost pricing and provide both guarantees of service and certainty of price in a statistically shared network.

The findings of this paper provide some suggestion of the type of products and services that will be made available. The widespread preference for price certainty guarantees the continued demand for services provided at flat rates. Yet demand for advanced services that require higher levels of bandwidth provided more reliably than best efforts will force customers of such advanced services to accept some uncertainty in price. This may have adverse effects on diffusion and universal service for broadband services.

The connectivity products thus will either require a higher flat rate to cover the uncertainty experience by the network owner (who must cover marginal cost) or embed the uncertainty in the tariff.

The long stated 'truism' that if video is affordable then voice is free has been proven to be untrue, at least for the moment. Broadband services allow the download of music and videos to the home while the revenues for voice services remain significant. Our analysis suggests that there is an underlying theoretical reason why video can be affordable and voice can be more costly to the consumer, as long as the video can be downloaded over a link with bandwidth uncertainty while voice demands real-time service. In fact, this implies that a premium price for voice is not only sustainable but required as voice requires greater certainty in bandwidth.

Further work obviously requires relaxing the heroic assumptions underlying this brief analysis, and more stringent statistical analysis of DSL pricing based on the certainty of bandwidth. In addition, the higher degree of certainty provided with frame relay should be analyzed with some rigor.